\documentclass[twocolumn,twoside,slac_two]{revtex4}
\usepackage{graphicx}
\usepackage{fancyhdr}
\begin{document}

\title{Luminosity Evolution of Gamma-ray Pulsars}

%

\author{Kouichi Hirotani}
\affiliation{Theoretical Institute for
       Advanced Research in Astrophysics (TIARA),
       Academia Sinica, Institute of Astronomy and Astrophysics (ASIAA),
       PO Box 23-141, Taipei, Taiwan}
%

\begin{abstract}
We investigate the electrodynamic structure of
a pulsar outer-magnetospheric particle accelerator
and the resultant gamma-ray emission.
By considering the condition for the
accelerator to be self-sustained,
we derive how the trans-magnetic-field thickness of the accelerator
evolves with the pulsar age.
It is found that the thickness is small but increases steadily
if the neutron-star envelope is contaminated by sufficient light elements.
For such a light element envelope,
the gamma-ray luminosity of the accelerator 
is kept approximately constant as a function of age
in the initial ten thousand years,
forming the lower bound of the observed distribution
of the gamma-ray luminosity of rotation-powered pulsars.
If the envelope consists of only heavy elements, on the other hand,
the thickness is greater but increases less rapidly
than what a light element envelope has.
For such a heavy element envelope,
the gamma-ray luminosity decreases relatively rapidly,
forming the upper bound of the observed distribution.
The gamma-ray luminosity of a general pulsar 
resides between these two extreme cases, 
reflecting the envelope composition and the magnetic inclination angle
with respect to the rotation axis.
The cutoff energy of the primary curvature emission is
regulated below several GeV even for young pulsars,
because the gap thickness, and hence the acceleration electric field
is suppressed by the polarization of the produced pairs.

\end{abstract}

\maketitle

\thispagestyle{fancy}


\section{Introduction}

The Large Area Telescope aboard {\it Fermi Gamma-ray Space Telescope} 
(\cite{atwood09}) has proved remarkably successful 
at discovering rotation-powered pulsars emitting photons above
0.1~GeV.
Thanks to its superb sensitivity,
the number of gamma-ray pulsars has increased from
six in Compton Gamma Ray Observatory era (\cite{thomp04})
to more than one hundred (\cite{nolan12}). 
Plotting their best estimate of
the gamma-ray luminosity, $L_\gamma$, 
against the spin-down luminosity,
$L_{\rm spin}=4\pi^2 I \dot{P} P^{-3}$,
\cite{abd10} found the important relation
that $L_\gamma$ is approximately proportional to 
$L_{\rm spin}{}^{0.5}$ (with a large scatter), 
where $I$ refers to the neutron-star (NS) moment of inertia,
$P$ the NS rotational period, and 
$\dot{P}$ its temporal derivative.
However, it is unclear why this relationship arises, 
in spite of its potential importance to discriminate
pulsar emission models such as 
the polar-cap model (\cite{harding78,daugherty82,dermer94}),
the outer-gap model (\cite{chiang92,romani96,zhang97,tak06,hiro08,wangY11}),
the pair-starved polar-cap model (\cite{vent09})
(see also \cite{yuki12} for the possible co-existence of such models),
and the emission model 
from the wind zone (\cite{petri10,bai10a,bai10b,aha12}).

Recent gamma-ray observations 
suggest that the pulsed gamma-rays are emitted from the
higher altitudes of a pulsar magnetosphere.
This is because the observed light curves (\cite{abd10}) favor 
fan-like emission geometry, 
which scan over a large fraction of the celestial sphere,
and because the Crab pulsar shows pulsed photons 
near and above 100~GeV (\cite{aliu11,ale11a,ale11b}),
which rules out an emission from the lower altitudes,
where strong magnetic absorption takes place for $\gamma$-rays
above $10$~GeV.
Consequently, higher-altitude emission models
such as the outer-gap model (\cite{cheng86a,cheng86b}),
the high-altitude slot-gap model (\cite{musl04}),
or the pair-starved polar-cap model (\cite{vent09}),
gathered attention.
It is noteworthy that the outer-gap model is presently 
the only higher-altitude emission model that is solvable 
from the basic equations self-consistently (\cite{hiro11a}). 
In the present paper, therefore, 
we focus on the outer-gap model and derive the
observed relationship $L_\gamma \propto L_{\rm spin}{}^{0.5}$
both analytically and numerically.

We schematically depict 
the pulsar outer-magnetospheric accelerator (i.e., the outer gap)
in figure~1 of \cite{hiro13} 
As the NS rotates, there appears the light cylinder,
within which plasmas can co-rotate with the magnetosphere.
The magnetic field lines that become tangential to the
light cylinder at the light cylinder radius,
$\varpi_{\rm LC}=cP/2\pi$, 
are called the last-open magnetic field lines,
where $c$ refers to the speed of light.
Pairs are produced via photon-photon pair production 
mostly near the null-charge surface 
and quickly polarized by 
the magnetic-field-aligned electric field, $E_\parallel$, in the gap. 
In this paper, we assume that the
rotation and magnetic axes reside in the same hemisphere
to obtain $E_\parallel > 0$,
which accelerates positrons ($e^+$'s) outwards while
electrons ($e^-$'s) inwards.
These ultra-relativistic particles 
have Lorentz factors, $\gamma \sim 10^{7.5}$,
to emit photons efficiently by the curvature process.


\section{Analytical examination of outer-gap luminosity}
\label{sec:analytical}
%
%
In this section, we analytically derive 
how the gamma-ray luminosity of an outer gap evolves with time.
In the outer magnetosphere, only the dipole component remains
in the magnetic field;
thus, the inhomogeneous part of the Maxwell equation 
(i.e., the Poisson equation for the electro-static potential)
gives the magnetic-field-aligned electric field \citep{hiro08}, 
\begin{equation}
  E_\parallel \approx \frac{\mu}{2\varpi_{\rm LC}^3} h_{\rm m}^2,
  \label{eq:app_Ell}
\end{equation}
where $\mu$ denotes the NS magnetic dipole moment,
and $h_{\rm m}$ the trans-magnetic-field thickness of the gap.
Since the Poisson equation is a second-order differential equation,
$E_\parallel$ is proportional to $h_{\rm m}{}^2$.
Electrons ($e^-$'s) and positrons ($e^+$'s) 
are created via photon-photon 
(and sometimes via magnetic) pair production, 
being subsequently polarized by $E_\parallel$ 
and accelerated in the opposite directions,
to finally attain the terminal Lorentz factor 
\begin{equation}
  \gamma = \left( \frac{3 \rho_{\rm c}^2}{2e} E_\parallel 
           \right)^{1/4},
  \label{eq:app_Lf}
\end{equation}
where $\rho_{\rm c}$ refers to the radius of curvature of
particle's motion in the three-dimensional magnetosphere,
$e$ the charge on the positron.
Photons are radiated by such ultra-relativistic
$e^\pm$'s via curvature process with characteristic energy,
\begin{equation}
  h \nu_{\rm c}= \frac32 \hbar c \frac{\gamma^3}{\rho_{\rm c}},
  \label{eq:app_Eg}
\end{equation}
where $h$ denotes the Planck constant,
$\hbar \equiv h/2\pi$.
Once $h_{\rm m}$ is obtained, 
we can readily compute the $\gamma$-ray luminosity
of curvature radiation from an outer gap by \citep{hiro08}
\begin{equation}
  L_\gamma \approx 2.36 (\nu F_\nu)_{\rm peak} \times 4\pi d^2 f_\Omega
          \approx 1.23 f_\Omega h_{\rm m}^3 
                  \frac{\mu^2 \Omega^4}{c^3},
  \label{eq:Lg_0}
\end{equation}
where $f_\Omega$, 
which has been conventionally assumed to be approximately unity, 
refers to the flux correction factor \citep{romani10},
and $\Omega=2\pi/P$ the rotation angular frequency of the NS.
Here, it is assumed that the current density flowing in the gap
is comparable to the Goldreich-Julain value \citep{gol69}.
The last factor, $\mu^2 \Omega^4/c^3$ is proportional to the spin-down 
luminosity, $L_{\rm spin}$.
Therefore, the evolution law, $L_\gamma \propto L_{\rm spin}^{0.5}$, 
is crucially governed by the evolution of $h_{\rm m}$
as a function of the NS age, $t$.

The evolution of $h_{\rm m}$ is essentially controlled by
the photon-photon pair production in the pulsar magnetosphere.
To analytically examine the pair production,
we assume the static dipole magnetic field configuration
for simplicity,
and consider the plane on which both the rotational and magnetic
axes reside. 
On this two-dimensional latitudinal plane, 
the last-open field line intersects the NS surface
at magnetic co-latitudinal angle $\theta_\ast^{\rm max}$
(measured from the magnetic dipole axis) that satisfies
\begin{equation}
  \frac{\sin^2\theta_\ast^{\rm max}}{r_\ast}
  = \frac{\sin^2(\theta_{\rm LC}-\alpha)}
         {\varpi_{\rm LC}/\sin\theta_{\rm LC}},
\end{equation}
where $r_\ast$ denotes the NS radius,
$\theta_{\rm LC}$ the angle 
(measured from the rotation axis)
of the point where the last-open field line becomes
tangential to the light cylinder,
and $\alpha$ the inclination angle of the dipole magnetic axis
with respect to the rotation axis.
A magnetic field line can be specified by the magnetic
co-latitude (measured from the dipole axis), 
$\theta_\ast$,
where it intersects the stellar surface.
A magnetic field line does not close within the light cylinder
(i.e., open to large distances)
if $0<\theta_\ast<\theta_\ast^{\rm max}$.
Thus, the last-open field lines, $\theta_\ast = \theta_\ast^{\rm max}$, 
corresponds to the {\it lower} boundary, 
which forms a surface in a three-dimensional magnetosphere, 
of the outer gap.

Let us assume that the gap {\it upper} boundary 
coincides with the magnetic field lines that are specified by
$\theta_\ast= (1-h_{\rm m})\theta_\ast^{\rm max}$.
Numerical examinations show
that $h_{\rm m}$, indeed, changes as a function of the distance
along the field line and the magnetic azimuthal angle
(measured counter-clockwise around the dipole axis).
Nevertheless, except for young pulsars like the Crab pulsar,
an assumption of a spatially constant $h_{\rm m}$ gives
a relatively good estimate.
Thus, for an analytical purpose,
we adopt a constant $h_{\rm m}$ in this analytical examination.
In this case, we can specify
the middle-latitude field line by the magnetic co-latitude
$\theta_\ast= (1-h_{\rm m}/2)\theta_\ast^{\rm max}$.
Screening of $E_\parallel$ due to the polarization of the
produced pairs,
takes place mostly in the lower altitudes.
It is, therefore, appropriate to evaluate the screening of $E_\parallel$
around the point ($r_0$,$\theta_0$)
where the null-charge surface intersects the middle-latitude field line.

An inwardly migrating electron
(or an outwardly migrating positron)
emits photons inwards (or outwards),
which propagate the typical distance $l_1$ (or $l_2$)
before escaping from the gap.
Denoting the cross section of the inward (or outward) horizontal line
from the point ($r_0$,$\theta_0$) and the
upper boundary as ($r_1$,$\theta_1$)
(or as ($r_2$,$\theta_2$)),
and noting 
$r_0 \cos\theta_0=r_1\cos\theta_1=r_2\cos\theta_2$, 
we obtain
\begin{equation}
  l_1= r_0 \cos\theta_0 (\tan\theta_0 -\tan\theta_1),
  \label{eq:S_l1}
\end{equation}
\begin{equation}
  l_2= r_0 \cos\theta_0 (\tan\theta_2 -\tan\theta_0).
  \label{eq:S_l2}
\end{equation}
Along the upper-boundary field line, we obtain
\begin{equation}
    \frac{\sin^2(\theta_1-\alpha)}{r_1}
  = \frac{\sin^2(\theta_2-\alpha)}{r_2}
  = \frac{\sin^2[(1-h_{\rm m})\theta_\ast^{\rm max}]}
         {r_\ast},
\end{equation}
whereas along the middle-latitude field line, we obtain
\begin{equation}
  \frac{\sin^2(\theta_0-\alpha)}{r_0}
  = \frac{\sin^2[(1-h_{\rm m}/2)\theta_\ast^{\rm max}]}
         {r_\ast}.
\end{equation}
Combining these two equations,
and noting $\theta_\ast^{\rm max} \ll 1$,
we find that $\theta_1$ ($<\theta_0$) and 
$\theta_2$ ($>\theta_0$)
can be given by the solution $\theta$ that satisfies
\begin{equation} 
  \cos\theta\sin^2(\theta-\alpha)
  = \left(\frac{1-h_{\rm m}}{1-h_{\rm m}/2}\right)^2
    \cos\theta_0 \sin^2(\theta_0-\alpha),
  \label{eq:S_th1}
\end{equation}
where $\theta_0$ is given by
\begin{equation} 
  \tan\theta_0
  = \frac12 
    \left( 3\tan\alpha+\sqrt{9\tan^2\alpha+8} \right). 
  \label{eq:S_th0}
\end{equation}
Thus, if we specify $\alpha$, 
we can solve $\theta=\theta_1$ and $\theta=\theta_2$ 
as a function of $h_{\rm m}$ by equation~(\ref{eq:S_th1}).
Substituting these $\theta_1$ and $\theta_2$ into
equations~(\ref{eq:S_l1}) and (\ref{eq:S_l2}),
we obtain $l_1$ and $l_2$,
where $r_0$ depends on $P$.

If $h_{\rm m} \ll 1$, we can expand the left-hand side of
equation~(\ref{eq:S_th1}) around
$\theta=\theta_0$,
where $\theta=\theta_1$ for inward 
(or $\theta=\theta_2$ for outward) $\gamma$-rays
to find $\theta_2-\theta_0= \theta_0-\theta_1 \propto \sqrt{h_{\rm m}}$.
That is, the leading terms in the expansion vanish and we obtain
$l_1= l_2$ from the next-order terms,
which are quadratic to $\theta-\theta_0$.
Assuming $L_{\rm X}\propto t^{-\beta}$,
where $\beta \approx 0.48$ is appropriate
for $t<10^4$ years for a light-element-envelope NS
and for $t<10^5$ years for a heavy-element NS,
we find
$h_{\rm m} \propto P^{5/6} \mu^{-1/6} t^{\beta/2}$,
and hence $L_\gamma \propto P^{-3/2} \mu^{3/2} t^{3\beta/2}$.
Since the dipole radiation formula gives
$P \propto \mu t^{1/2}$, we obtain
$L_\gamma \propto \mu^0 t^{3(\beta-1/2)/2} \propto t^{-0.03}$.
Thus, when the gap is very thin,
which is expected for a light-element-envelope NS,
$L_\gamma$ little evolves with the pulsar age, $t$.

However, if $h_{\rm m}>0.2$, say,
the rapidly expanding magnetic flux tube 
gives asymmetric solution, $l_2 > l_1$.
That is, the third and higher order terms in the expansion contribute
significantly compared to the quadratic terms.
Thus, we must solve equation~(\ref{eq:S_th1})
for $\theta$ ($=\theta_1$ or $\theta_2$)
without assuming $\vert \theta-\theta_0 \vert \ll 1$, in general.

Let us now consider the condition for a gap to be self-sustained.
A single ingoing $e^-$ or an outgoing $e^+$ emits
\begin{equation}
  (N_\gamma)_1= e E_\parallel l_1 / (h \nu_{\rm c})
\end{equation}
or
\begin{equation}
  (N_\gamma)_2= e E_\parallel l_2 / (h \nu_{\rm c})
\end{equation}
photons while running the typical distance $l_1$ or $l_2$, respectively.
Such photons materialize as pairs with probability
\begin{equation}
  \tau_1= l_1 F_1 \sigma_1 / c
\end{equation}
or 
\begin{equation}
  \tau_2= l_2 F_2 \sigma_2 / c,
\end{equation}
where $F_1$ and $F_2$ denote the X-ray flux 
inside and outside of ($r_0$,$\theta_0$), respectively; 
$\sigma_1$ and $\sigma_1$ are the pair-production cross section
for inward and outward photons, respectively.
Thus, a single $e^-$ or $e^+$ cascades into 
\begin{equation}
  (N_\gamma)_1 \tau_1= \frac{e E_\parallel}{h \nu_{\rm c}}
                      \frac{F_1}{c}
                      l_1{}^2 \sigma_1
  \label{eq:S_N1}
\end{equation}
pairs or into
\begin{equation}
  (N_\gamma)_2 \tau_2= \frac{e E_\parallel}{h \nu_{\rm c}}
                      \frac{F_2}{c}
                      l_2{}^2 \sigma_2
  \label{eq:S_N2}
\end{equation}
pairs within the gap.
That is, a single inward-migrating $e^-$ cascades into 
pairs with multiplicity $(N_\gamma)_1 \tau_1$.
Such produced pairs are polarized by $E_\parallel$.
Each returning, outward-migrating $e^+$ cascades into
pairs with multiplicity $(N_\gamma)_2 \tau_2$ in outer magnetosphere.
As a result, a single inward $e^-$ cascades eventually
into $(N_\gamma)_1 \tau_1 \cdot (N_\gamma)_2 \tau_2$ inward $e^-$'s,
which should become unity for the gap to be self-sustained.
Therefore, in a stationary gap,
the gap thickness $h_{\rm m}$ is automatically regulated
so that the gap closure condition,
\begin{equation}
  (N_\gamma)_1 \tau_1 \cdot (N_\gamma)_2 \tau_2 = 1,
  \label{eq:S_stationary}
\end{equation}
may be satisfied.

Approximately speaking, a single, inward-migrating
$e^-$ emits $(N_\gamma)_1 \sim 10^4$ curvature photons,
a portion of which head-on collide the surface X-ray photons
to materialize as pairs with probability
$\tau_1 \sim 10^{-3}$ within the gap.  
Thus, a single $e^-$ cascades into 
$(N_\gamma)_1 \tau_1 \sim 10$ pairs in the gap.
Each produced $e^+$ return outwards to emit 
$(N_\gamma)_2 \sim 10^5$ photons,
which materialize as pairs with probability
$\tau_2 \sim 10^{-6}$ 
by tail-on colliding with the surface X-rays.
In another word, $(N_\gamma)_1 \tau_1\sim 10$ 
holds regardless of the nature of the pair production process
(e.g., either photon-photon or magnetic process \citep{tak10})
in the lower altitudes,
because 
it is determined by the pair-production efficiency
in the outer magnetosphere
$(N_\gamma)_2 \tau_2\sim 0.1$,
which is always due to photon-photon pair production.

In general, 
$(N_\gamma)_1$, $\tau_1$, $(N_\gamma)_2$, $\tau_2$
are expressed in terms of 
$h_{\rm m}$, $P$, $\mu$, $T$, and $\alpha$,
where $T$ denotes the NS surface temperature.
Note that we can solve $P=2\pi/\Omega$ as a function of 
the NS age, $t$,
from the spin-down law.
Thus, specifying $\alpha$ and the cooling curve, $T=T(t)$,
we can solve $h_{\rm m}$ as a function of $t$
from the gap closure condition,
$(N_\gamma)_1 \tau_1 (N_\gamma)_2 \tau_2 = 1$.
Note also that the spin-down law
readily gives the spin-down luminosity,
$L_{\rm spin} \propto \dot{P} P^{-3}$,
as a function of $t$, 
once $P=P(t,\alpha)$ is solved.
On these grounds,
$L_\gamma$ can be related to $L_{\rm spin}$
with an intermediate parameter $t$,
if we specify the cooling curve and the spin-down law.

Substituting equations~(\ref{eq:S_N1}) and (\ref{eq:S_N2})
into (\ref{eq:S_stationary}), 
we obtain 
\begin{equation}
  \frac{e E_\parallel}{h\nu_{\rm c}}
  \frac{\sqrt{F_1 \sigma_1 F_2 \sigma_2}}{c}
  l_1 l_2 = 1,
  \label{eq:S_master}
\end{equation}
where 
\begin{equation}
  F_i \sigma_i
  = \pi (1-\mu_i) \left(\frac{r_\ast}{r_i}\right)^2
    \int_{\nu_{{\rm th},i}}^\infty \frac{B_\nu(T)}{h\nu} 
                      \sigma_{\rm P}(\nu,\nu_\gamma,\mu_i)
\end{equation}
with $i=1,2$; $\nu_\gamma$ denotes
the $\gamma$-ray frequency, and $B_\nu(T)$ the Planck function.
We have to integrate over the soft photon frequency
$\nu$ above the threshold energy
\begin{equation}
  h\nu_{{\rm th},i}= \frac{2(m_{\rm e}c^2)^2}
               {(1-\mu_i)h\nu_\gamma},
\end{equation}
where $m_{\rm e}c^2$ refers to the rest-mass energy of the electron.
The cosine of the collision angle $\mu_i$ becomes
$1-\mu_1 = 1 - \sin\theta_0$ for outward 
(or $1-\mu_2 = 1+\sin\theta_0$ for inward) $\gamma$-rays.
That is, collisions take place head-on (or tail-on)
for inward (or outward) $\gamma$-rays.
The total cross section is given by
\begin{equation}
  \sigma_{\rm P}
  =\frac{3}{16} \sigma_{\rm T} 
   (1-v^2)
   \left[ (3-v^4) \ln\frac{1+v}{1-v} -2v(2-v^2) \right],
\end{equation}
where $\sigma_{\rm T}$ denotes the Thomson cross section and
\begin{equation}
  v \equiv \sqrt{1-\frac{2}{1-\mu_i}
                   \frac{(m_{\rm e}c^2)^2}{h\nu h\nu_\gamma}}.
\end{equation}

Pair production takes place when the $\gamma$-rays collide with the
surface X-rays in the Wien regime,
that is, at $h\nu \gg kT$.
An accurate evaluation of $\sigma_2$ requires
a careful treatment of the collision geometry,
because the threshold energy, $h\nu_{{\rm th},2}$,
strongly depends on the tiny collision angles.
In the numerical method (next section),
the pair-production absorption coefficient is explicitly computed
at each point in the three-dimensional pulsar magnetosphere.
However, in this section, for analytical purpose,
we simply adopt the empirical relation,
\begin{equation}
  \sqrt{F_1 \sigma_1 F_2 \sigma_2}
  = \epsilon \sqrt{1-\mu_1} \sigma_{\rm T} F_{\rm X},
  \label{eq:L1L2}
\end{equation}
where $\epsilon=0.004$, $0.01$, and $0.038$ for
$\alpha=45^\circ$, $60^\circ$, and $75^\circ$, respectively;
$1-\mu_1 \approx 2$.
The X-ray flux is evaluated at ($r_0$,$\theta_0$) such that
\begin{equation}
  F_{\rm X} = \frac{L_{\rm X}}{2.82 kT}
             \frac{1}{4\pi r_0{}^2},
\end{equation}
where $L_{\rm X}$ refers to the luminosity of photon radiation 
from the the cooling NS surface.
For a smaller $\alpha$, the point ($r_2$,$\theta_2$) is located
in the higher altitudes,
where the magnetic field lines begin to collimate along the 
rotation axis,
deviating from the static dipole configuration.
Thus, the collision angles near the light cylinder,
and hence $\sigma_2$ decreases with decreasing $\alpha$.
The explicit value of $\epsilon$ can be computed only numerically,
solving the photon specific intensity from infrared to $\gamma$-ray
energies in the three-dimensional pulsar magnetosphere.

The last factor, $l_1 l_2$, in the left-hand side of
equation~(\ref{eq:S_master}) is given by
\begin{equation} 
  l_1 l_2 = r_0{}^2 \cos^2\theta_0
           (\tan\theta_0-\tan\theta_1)
           (\tan\theta_2-\tan\theta_0)
  \label{eq:l1l2}
\end{equation}
Thus, equation~(\ref{eq:S_master}) gives
\begin{eqnarray}
  &&
  \frac{e E_\parallel}{h\nu_{\rm c}}
  \frac{L_{\rm X}/c}{2.82 kT}
  \epsilon \sqrt{1-\mu_1} \sigma_{\rm T}
  \nonumber\\
  && \times
  \cos^2\theta_0
           (\tan\theta_0-\tan\theta_1)
           (\tan\theta_2-\tan\theta_0)
  = 1,
  \label{eq:S_master2}
\end{eqnarray}
where the $r_0$ dependence vanishes.
Substituting equations~(\ref{eq:app_Ell}),
(\ref{eq:app_Lf}), (\ref{eq:app_Eg}) into (\ref{eq:S_master2}),
we can solve $h_{\rm m}$ as a function of
$L_{\rm X}/kT$, $P$, and $\mu$.

To describe the evolution of $P=P(t)=2\pi/\Omega(t)$,
we adopt in this paper
\begin{equation}
  -I \Omega \dot{\Omega}= C \frac{\mu^2 \Omega^4}{c^3}
\end{equation}
where $C=(2/3)\sin^2\alpha$ for a magnetic dipole braking,
while $C=1+\sin^2\alpha$ for a force-free braking \citep{spit06}.
Assuming a magnetic dipole braking,
we obtain
\begin{equation}
  P= 39.2 \mbox{ms} \mu_{30} I_{45}^{-1/2} (t/10^3 \mbox{years})^{1/2},
\end{equation}
where $\mu_{30} \equiv \mu / (10^{30}\,\mbox{G cm}^3)$
and $I_{45} \equiv I/(10^{45}\,\mbox{g cm}^2)$.
Thus, if we specify a cooling scenario, $T=T(t)$,
equation~(\ref{eq:S_master2}) gives
$h_{\rm m}$ as a function of $t$.
Note that the $\alpha$ dependence of the spin-down law
is not essential for the present purpose;
thus, $C=2/3$ is simply adopted.
Once $h_{\rm m}=h_{\rm m}(t)$ is obtained,
equation~(\ref{eq:Lg_0}) readily gives 
$L_\gamma$ as a function of $t$,
and hence of $L_{\rm spin}$.
It is worth noting that the heated polar-cap emission
is relatively weak compared to the cooling NS emission,
except for millisecond or middle-aged pulsars.

We adopt the minimal cooling scenario \citep{pag04}
Within the minimal cooling scenario, 
the cooling history of a NS substantially depends
on the composition of the envelope.
We adopt the cooling curves given in \citet{pag04}
and consider the two extreme cases:
light element and heavy element envelopes.


We present the solved $h_{\rm m}$ for a light and a heavy
element envelope in figure~2 in \cite{hiro13}.
It follows that the gap becomes thinner for
a light element case than for the heavy element cases.
This is because the more luminous photon field of 
a light element envelope leads to a copious pair production,
which prevents the gap to expand in the trans-field direction.
As a result, the predicted $L_\gamma$ 
becomes less luminous for a light element envelope
than a heavy one. 

In figure~\ref{fig:LgLsp_2},
we present the analytical results of $L_\gamma$ versus
$L_{\rm spin}$
as the dotted (or dashed) curve for a light (or a heavy) 
element envelope.
As the pulsar spins down, $L_\gamma$ evolves leftwards.
It is interesting to note that $L_\gamma$ little
evolves for a light element envelope,
as explained in \cite{hiro13}.

\section{Numerical examination of outer-gap electrodynamics}
\label{sec:numerical}
Let us develop the analytical examination
and look deeper into a self-consistent solution by 
a numerical method.
To this end, we adopt the modern outer-gap model \citep{hiro11a}
and solve the set of Maxwell and Boltzmann equations
self-consistently
and compute $E_\parallel$, distribution functions of $e^\pm$'s,
and the photon specific intensity
at each point in the three-dimensional pulsar magnetosphere.
We consider not only the whole-surface, cooling NS emission
but also the heated polar-cap emission as the photon source of
photon-photon pair production in the numerical analysis.
The former emission component is given as a function of the 
pulsar age from the minimum cooling scenario, 
in the same manner as in the analytical examination,
while the latter emission component is solved 
consistently with the energy flux of the $e^-$'s falling on to
the pulsar polar-cap surface.

The numerical method is described in \cite{hiro13} in detail.
We solve the set of partial and ordinary 
differential equations
under the boundary conditions that $e^\pm$'s or $\gamma$-rays
do not penetrate into the gap from outside.
By this method, we can solve 
the acceleration electric field $E_\parallel$,
particle distribution functions $n_\pm$, and
the photon specific intensity $I_\nu$ (from $h\nu=0.005$~eV to $50$~TeV),
at each position in the three-dimensional magnetosphere of
arbitrary rotation-powered pulsars,
if we specify $P$, $\mu$, $\alpha$, and $kT$.
We adopt the minimum cooling scenario in the same manner as in 
\S~\ref{sec:analytical}.

\begin{figure*}[t]
\centering
\includegraphics[width=135mm]{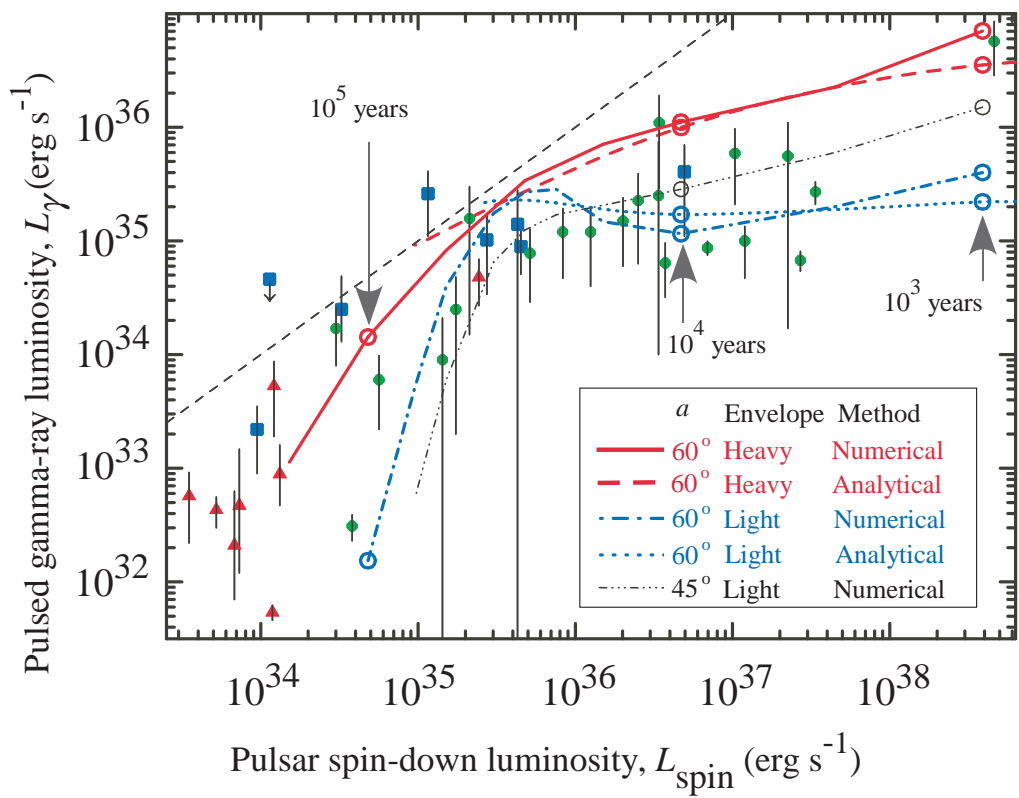}
\caption{
Evolution of the outer-gap luminosity
as a function of the neutron-star spin-down luminosity.
Analytical results are plotted as dotted and dashed curves,
while numerical ones as dash-dotted, solid, and dash-dot-dot-dot ones.
A light or a heavy element envelope is assumed, as indicated in the box.
For comparison, the case of $\alpha=45^\circ$ is also depicted.
The green filled circles designates the normal gamma-ray pulsars,
while the blue filled squares do those detected 
by the gamma-ray blind search technique.
} \label{fig:LgLsp_2}
\end{figure*}

In figure~\ref{fig:LgLsp_2},
we plot the result of $L_\gamma$ as a function of 
$L_{\rm spin}$
as the dash-dotted (or solid) curve
for a light (or a heavy) element envelope,
where $\mu_{30}=3.2$ is adopted in the same manner as in the
analytical examination.
It follows that these numerical solutions 
are consistent with the analytical ones,
and that $L_\gamma$ decreases slowly until 
$10^{4.5}$ years.
The physical reason why $L_\gamma$ increases 
with decreasing $L_{\rm spin}$ at $t>10^4$~years for a light element
envelope,
is the same as described at the end of \S~\ref{sec:analytical}.
A realistic NS will have an envelope composition
between the two extreme cases, light and heavy elements.
Thus, the actual $L_\gamma$'s will distribute between the 
red solid (or dashed) and the blue dash-dotted (or dotted) curves.
However, after $L_\gamma$ approaches $L_{\rm spin}$
(thin dashed straight line; 
 see \citet{wangR11} for the death line argument),
the outer gap survives only along the limited magnetic field lines
in the trailing side of the rotating magnetosphere
because of a less efficient pair production;
as a result, $L_\gamma$ rapidly decreases with decreasing $L_{\rm spin}$.
For a smaller $\alpha$, even for a light element envelope,
$L_\gamma$ monotonically decreases 
as the dash-dot-dot-dot curve shows,
because the gap is located in the higher altitudes,
and because the less efficient pair production there prevents
the produced electric current to increase with decreasing age around
$t \sim 10^{4.5}$~years.

\section{Discussion}
\label{sec:discussion}
To sum up, 
a light element envelope approximately corresponds to the
lower bound of the (observationally inferred) gamma-ray luminosity 
of rotation-powered pulsars,
whereas a heavy element one to the upper bound.
The scatter of the intrinsic gamma-ray luminosity 
is physically determined by
the magnetic inclination angle, $\alpha$, and the envelope composition.
The cutoff energy of the primary curvature emission is
kept below several GeV even for young pulsars,
because the gap trans-field thickness, and hence the acceleration 
electric field, is suppressed by the polarization of the
produced pairs in the lower altitudes.

To convert the observed $\gamma$-ray flux into luminosity, $L_\gamma$,
one has conventionally assumed $f_\Omega=1$.
For example, the error bars of the observational data points 
in figure~\ref{fig:LgLsp_2},
do not contain any uncertainties incurred by $f_\Omega$.
Nevertheless, if $\alpha$ and $\zeta$ can be constrained,
we can estimate $L_\gamma$ more accurately,
by applying the present quantitative outer-gap calculations.
It is noteworthy that $L_\gamma$'s given in 
figure~\ref{fig:LgLsp_2} 
little depend on the NS magnetic moment, $\mu$.
This is particularly true for a light element case,
which has $h_{\rm m} \ll 1$,
by the reason described after equation~(\ref{eq:S_th0}).
What is more, 
with an additional determination of $d$ 
(e.g., by parallax observations),
we can infer the composition of individual NS envelopes,
by using the constrained flux correction factor,
$f_\Omega$ (fig.~6 in \cite{hiro13}).
We hope to address such a question as the determination of
$\alpha$ and $\zeta$, and hence $f_\Omega$, for individual pulsars,
by making an \lq atlas' of the pulse profiles and phase-resolved spectra
that are solved from the basic equations
in a wide parameter space of $P$, $\mu$, $T$, $\alpha$, and $\zeta$,
and by comparing the atlas with the observations.

\bigskip 
\begin{acknowledgments}
The author is indebted to Dr. A.~K. Harding for valuable discussion
on the results.
He also thanks ASPEN Center for Physics 
for providing precious opportunity to debate the main topic
of this letter.
This work is partly supported by 
the Formosa Program between National Science Council  
in Taiwan and Consejo Superior de Investigaciones Cientificas
in Spain administered through grant number 
NSC100-2923-M-007-001-MY3.
\end{acknowledgments}

\bigskip 

\begin{thebibliography}{9}   


\bibitem[Abdo et al.(2010)]{abd10} 
  Abdo, A. A. et al., 2010, ApJS, 187, 460

\bibitem[Aharonian et al.(2012)]{aha12} 
  Aharonian, F. A., Bogovalov, S. V. \&  Khangulyan, D.
  2012, Nature 482, 507


\bibitem[Aleksi\'c et al.(2011a)]{ale11a} 
  Aleksi\'c, J. {\it et al.} 
  2011a, ApJ 742, 43

\bibitem[Aleksi\'c et al.(2011b)]{ale11b} 
  Aleksi\'c, J., {\it et al.} 
  2011b, A\&Ap 540, 69

\bibitem[Aliu et al.(2011)]{aliu11} 
  Aliu, E. Arlen, T., Aune, T., {\it et al.}
  2011, Science 334, 69

\bibitem[Atwood et al.(2009)]{atwood09} 
  Atwood, W. B. et al.
  2009, ApJ 697, 1071

\bibitem[Bai \& Spitkovski(2010a)]{bai10a} 
  Bai, X. N. \& Spitkovski, A.
  2010a, ApJ 715, 1270

\bibitem[Bai \& Spitkovski(2010b)]{bai10b} 
  Bai, X. N. \& Spitkovski, A.
  2010b, ApJ 715, 1282

%

%
\bibitem[Cheng et al.(1986a)]{cheng86a} 
  Cheng, K. S., Ho, C. \&  Ruderman, M.
  1986a, ApJ 300, 500

\bibitem[Cheng et al.(1986b)]{cheng86b} 
  Cheng, K. S., Ho, C. \&  Ruderman, M.
  1986b, ApJ 300, 522



\bibitem[Chiang \& Romani(1992)]{chiang92} 
  Chiang, J. \& Romani, R. W.n
  1992, ApJ 400, 629

\bibitem[Daugherty \& Harding(1982)]{daugherty82} 
  Daugherty, J. K. \& Harding, A. K.
  1982, ApJ 252, 337

%
\bibitem[Dermer(1994)]{dermer94} 
  Dermer, C. D. \& Sturner, S. J.
  1994, ApJ 420, L75

\bibitem[Goldreich, \&  Julian(1969)]{gol69} 
  Goldreich, P. \&  Julian, W. H.
  1969, ApJ 157, 869

\bibitem[Harding et al.(1978)]{harding78} 
  Harding, A. K., Tademaru, E. \& Esposito, L. S.
  1978, ApJ 225, 226



\bibitem[Hirotani(2008)]{hiro08}
  Hirotani, K.
  2008, ApJ 688, L25

\bibitem[Hirotani(2011a)]{hiro11a}
  Hirotani, K.
  2011a, 
  The first session of the Sant Cugat Forum Astrophysics
  (eds Rea, N. \& Torres, D.~F.) p. 117
  (Springer, Berlin)


\bibitem[Hirotani(2013)]{hiro13}
  Hirotani, K.
  2013, ApJ in press.








\bibitem[Muslimov \& Harding(2004)]{musl04} 
  Muslimov, A. \& Harding, A. K.
  2004, ApJ 606, 1143

\bibitem[Nolan(2012)]{nolan12} 
  Nolan, P. {\it et al.} 
  {\it Fermi} Large Area Telescope second source catalog
  {\it Astroph. J. Suppl.} {\bf 199}, 31 (2012).

\bibitem[Page et al.(2004)]{pag04} 
  Page, D., Lattimer, J.~M., Prakash, M. \& Steiner A.~W.
  2004, ApJS 155, 623

\bibitem[Petri(2011)]{petri10} 
  Petri, J.
  2011, MNRAS 412, 1870

\bibitem[Romani(1996)]{romani96} 
  Romani, R. W.
  1996, ApJ 470, 469

\bibitem[Romani, R. \& Watters(2010)]{romani10} %
  Romani, R. \& Watters, K. P.
  2010, ApJ 714, 810


\bibitem[Spitkovski(2006)]{spit06} %
  Spitkovsky, A.
  2006, ApJ 648, L51



\bibitem[Takata et al.(2004)]{tak06} %
  Takata, J., Shibata, S., Hirotani, K., and Chang, H.-K.
  2006, MNRAS 366, 1310


\bibitem[Takata et al.(2010)]{tak10} %
  Takata, J., Wang, Y. \& Cheng, K. S.
  2010, ApJ 715, 1318


\bibitem[Thompson(2004)]{thomp04} 
  Thompson, D. J. 
  in {\it Cosmic Gamma-Ray Sources}
  (eds Cheng, K.~S. \& Romero, G.~E.) 149
  (Astrophys. Space Sci. Lib. 304,
   Dordrecht, Kluwer, 2004).

\bibitem[Venter et al.(2009)]{vent09} 
  Venter, C., Harding, A. K. \& Guillemot, L.
  2009, ApJ 707, 800

\bibitem[Wang \& Hirotani(2011)]{wangR11}
  Wang, R.~B. \& Hirotani, K.
  2011, ApJ 736, 127

\bibitem[Wang et al.(2011)]{wangY11} 
  Wang, Y., Takata, J. \& Cheng, K. S.
  2011, MNRAS 414, 2664

\bibitem[Yuki \& Shibata(2012)]{yuki12} 
  Yuki, S. \& Shibata, S.
  2012, PASJ 64, 43

\bibitem[Zhang \& Cheng(1997)]{zhang97} 
  Zhang, J. L. \& Cheng, K. S.
  1997, ApJ 487, 370



\end{thebibliography}

\end{document}